\documentclass{article}

\usepackage{arxiv}

\usepackage[utf8]{inputenc}   % allow utf-8 input
\usepackage[T1]{fontenc}      % use 8-bit T1 fonts
\usepackage{hyperref}         % hyperlinks
\usepackage{url}              % simple URL typesetting
\usepackage{booktabs}         % professional-quality tables
\usepackage{amsmath}
\usepackage{amssymb}
\usepackage{amsfonts}
\usepackage{graphicx}
\usepackage[round,authoryear]{natbib}
\usepackage{microtype}

\graphicspath{{./}}

\newcommand{\anon}[2]{#1}

\title{Measuring Curriculum--Labor Market Alignment at the Scale of a Program Portfolio}

\author{%
  \textbf{Sherzod Turaev}\,$^{1,*}$ \quad \textbf{Saja Aldabet}\,$^{1}$ \quad \textbf{Mary John}\,$^{2}$ \quad
  \textbf{Namya Musthafa}\,$^{3}$ \\[4pt]
  \textbf{Mamoun Awad}\,$^{1}$ \quad \textbf{Nazar Zaki}\,$^{1}$ \quad \textbf{Khaled Shuaib}\,$^{4}$ \\[6pt]
  \mdseries\normalsize $^{1}$Department of Computer Science \& Software Engineering,
  College of Information Technology,\\
  \mdseries\normalsize United Arab Emirates University, Al Ain, United Arab Emirates\\[2pt]
  \mdseries\normalsize $^{2}$Academic Support Department, Abu Dhabi Polytechnic,
  Abu Dhabi, United Arab Emirates\\[2pt]
  \mdseries\normalsize $^{3}$Department of Electrical and Communication Engineering,
  College of Engineering,\\
  \mdseries\normalsize United Arab Emirates University, Al Ain, United Arab Emirates\\[2pt]
  \mdseries\normalsize $^{4}$Department of Information Systems and Security,
  College of Information Technology,\\
  \mdseries\normalsize United Arab Emirates University, Al Ain, United Arab Emirates\\[4pt]
  \mdseries\normalsize $^{*}$Corresponding author: \texttt{sherzod@uaeu.ac.ae}%
}

\hypersetup{
  pdftitle={Measuring Curriculum-Labor Market Alignment at the Scale of a Program Portfolio},
  pdfauthor={Sherzod Turaev, Saja Aldabet, Mary John, Namya Musthafa, Mamoun Awad, Nazar Zaki, Khaled Shuaib},
  colorlinks=true, linkcolor=black, citecolor=black, urlcolor=blue
}

\begin{document}
\maketitle

\begin{abstract}
A college that offers several overlapping computing degrees implicitly assumes that its programs are differentiated in line with the way the labor market segments computing work and that, taken together, they prepare graduates for that market. Testing such an assumption is difficult, because the instruments available to curriculum committees, namely advisory boards, graduate tracer studies, and employer surveys, are slow, narrow, and hard to reproduce. In the current work, we apply one uniform, taxonomy-anchored alignment analysis across all five undergraduate programs of a College of Information Technology, comparing $1{,}922$ course learning outcomes against $103{,}349$ competencies extracted from a unified corpus of $5{,}186$ deduplicated job openings drawn from four boards. Every competency, on both sides, is obtained by a \emph{grounded} single-language-model procedure that copies it verbatim from the source and verifies it against the source, then assigned to one of eleven ESCO-aligned domains and a Bloom cognitive level, and the curricular supply is read not as a catalog but on a \emph{realized-attainment} basis that respects the credit-hour and elective constraints under which a student completes a degree. The extraction is validated blind by two independent faculty raters (domain $\kappa = 0.91$, Bloom level $\kappa = 0.86$) and the ESCO matching by a human-adjudicated gold set ($\kappa = 0.72$). The analysis yields four findings. The content gaps are systemic rather than program-specific, concentrated in systems, software engineering, security, and web development; the shared college core satisfies only about a third of the demanded competencies; the programs are well differentiated in disciplinary content yet homogeneous in where they fall short; and the curriculum is aligned roughly a full Bloom level below the market across the portfolio, most acutely in systems. We discuss the implications for program design, for curriculum governance, and for the practice of curriculum analytics.
\end{abstract}

\vspace{4pt}
\vspace{6pt}

\section{Introduction}

The relationship between what universities teach and what the labor market requires has long been a concern of higher-education policy, and within computing it has become acute, because the field's occupational categories multiply and its tools turn over faster than the multi-year cycles on which curricula are revised \citep{cc2020}. A computing program is, in effect, a standing premise that the competencies it cultivates will still command value by the time its graduates enter the workforce. Where a college offers several computing degrees at once, the premise is compounded by a second one, namely that the programs are differentiated from one another in a way that mirrors how the labor market itself segments computing roles, so that a graduate of each program is prepared for a distinct slice of the market rather than for an undifferentiated average of it.

The instruments on which institutions traditionally rely to examine these wagers, namely industry advisory boards, graduate tracer studies, and employer satisfaction surveys, are valuable but inherently slow, narrow in coverage, and difficult to reproduce across programs or over time. A decade of computational work has sought to supplement them by mining curricular and labor-market text directly, and recent advances in large language models have made it possible to extract structured competency records, anchored to standardized skill taxonomies, with a reliability that the earlier lexical and shallow-semantic methods could not attain. A reference framework for such analysis, together with a single-program proof of concept, has been established in prior work \anon{\citep{turaev2026framework, turaev2026pilot}}{(citations withheld for anonymous review)}; that work, however, diagnoses one program in isolation.

What has not been examined is the \emph{portfolio}, that is, the full set of programs a college offers, analyzed as a system against a common labor market with one uniform instrument. The portfolio is the level at which the governance questions actually arise, because it is the college, and not the individual course, that decides which competencies are guaranteed to every graduate through a shared core, which are reserved to program-specific majors, and how the programs are meant to differ from one another. A study confined to a single program cannot speak to whether a gap is peculiar to that program or shared across all of them, whether the common core carries its weight, or whether the programs are differentiated in line with the market they collectively serve.

These governance choices are neither incidental nor easily reversed. A college must decide which competencies every graduate will be guaranteed through the shared core and which will be localized in a particular major, how far the sibling programs may overlap before differentiation collapses into redundancy, and how to weigh the breadth that prepares a graduate for several adjacent roles against the depth that genuine specialization requires. Each such decision is taken once and then binds a cohort for the several years of a degree, and each is made, in current practice, largely on the basis of disciplinary tradition, accreditation requirements, and the periodic judgment of advisory boards, rather than on a measured reading of how the programs together meet the market they serve. The portfolio is, in this sense, the natural unit of curricular governance, and yet it is precisely the unit at which systematic evidence is scarcest.

A second consideration concerns what a curriculum delivers as opposed to what it lists. A published plan of study is an inventory of opportunities rather than a record of attainment, because the general-education and elective requirements that bulk out a catalog are satisfied by selecting a few courses from a much larger menu, so that no single graduate encounters more than a fraction of what the catalog displays. A supply-versus-demand comparison that counts every cataloged course equally therefore measures an idealized program that no student actually completes, and it systematically overstates the breadth a degree can guarantee. Separating what a program offers from what a graduate is assured of receiving is, we argue, a precondition for an honest alignment analysis, and it is the second thread the present study develops.

In the current research, we take the college as the unit of analysis and apply one uniform, taxonomy-anchored alignment instrument across all five undergraduate programs of a College of Information Technology, namely Computer Science, Computer Engineering, Data Science and Artificial Intelligence, Information Security, and Information Technology. We compare the competencies these programs develop against those demanded by a contemporaneous corpus of regional job advertisements, on a common frame of eleven domains aligned to the European Skills, Competences, Qualifications and Occupations (ESCO) taxonomy and five Bloom cognitive levels. One methodological choice distinguishes the analysis: rather than counting every course in the catalog equally, we read the curricular supply on a \emph{realized-attainment} basis that respects the credit-hour and elective constraints under which a student completes a degree, so that the comparison reflects what graduates actually attain rather than what the catalog nominally offers. We organize the study around four questions: whether the curriculum--market gaps are systemic across the portfolio or specific to particular programs (RQ1); how much of the market's demand the shared college core satisfies on its own (RQ2); whether the programs are differentiated in line with the market's segmentation of computing roles (RQ3); and whether the curriculum is pitched at the cognitive depth the market expects (RQ4).

The paper makes four contributions:
\begin{enumerate}
\item the first portfolio-scale alignment analysis of an entire college of computing programs, conducted with a single uniform instrument so that programs are directly comparable rather than studied in isolation, and built on a \emph{grounded} extraction whose every competency is copied verbatim from its source and verified against it, with the extraction and the ESCO matching independently validated;
\item a \emph{realized-attainment} scope model that weights the curricular supply by the credit-hour and elective constraints of each program, correcting the catalog-level over-counting that would otherwise distort the supply--demand comparison;
\item an empirical diagnosis of the College of Information Technology that establishes the gaps as systemic rather than program-specific, quantifies the limited reach of the shared core, distinguishes content differentiation from gap homogeneity across the programs, and measures a portfolio-wide cognitive-depth deficit; and
\item a reading of these results for curriculum governance, identifying where in the plan, and at what cognitive level, each gap could be closed.
\end{enumerate}

The remainder of the paper is organized as follows. The next section reviews related work in computing-education curriculum design and computational curriculum--labor alignment and situates the present study. We then recap the alignment framework and the competency formalism on which the analysis rests, before detailing the study design and method, including the corpora, the grounded extraction, the taxonomic and cognitive anchoring, and the scope model. The reliability of the extraction and of the ESCO matching is established next, after which the results are reported around the four research questions, followed by a discussion of the implications for program design and governance, the limitations of the study, and conclusions.

\section{Background and Related Work}

That graduate competencies and employer expectations diverge is among the more persistently documented concerns in higher education, and the divergence is sharpest in computing. Labor-market analyses report that a substantial fraction of workers' core skills will change within a few years and that employers identify the skills gap as a leading barrier to growth \citep{wef2025jobs}, while the automation of routine cognitive work continues to reshape which competencies retain value \citep{frey2017future}. Within computing and software engineering specifically, studies have repeatedly documented a gap between what programs cultivate and what industry expects, in both technical and professional competencies \citep{garousi2020closing, radermacher2013gaps}. The difficulty is structural rather than incidental: universities revise curricula on multi-year cycles, whereas the competencies the market rewards shift continuously, so that even a well-designed program drifts out of alignment between revisions.

Computing-education practice has long sought to manage this through outcome- and competency-oriented curriculum design. The ACM and IEEE Computer Society curricular guidelines, culminating in the competency-based CC2020 framework \citep{cc2020}, define what graduates should know and be able to do, and accreditation regimes require programs to specify measurable student outcomes and to demonstrate their attainment \citep{abet2024criteria}, with national qualifications frameworks situating each degree at a defined level of competence. These practices rest on a longer tradition of outcome-based education \citep{spady1994obe} and of constructive alignment between intended outcomes, learning activities, and assessment \citep{biggs2011teaching}, in which the cognitive demand of an outcome is conventionally expressed on the revised Bloom taxonomy \citep{anderson2001taxonomy}. Valuable as they are for internal coherence and disciplinary conformance, these instruments specify what a program intends; they do not, on their own, measure or track how well those intentions match a labor market that changes faster than the curricular cycle.

The discipline does not, moreover, present itself as a single curriculum. The ACM and IEEE Computer Society overview frames computing as a family of related but distinct disciplines, each governed by its own curricular volume, so that separate guidelines now exist for computer science \citep{cs2023}, computer engineering \citep{ce2016}, information technology \citep{it2017}, cybersecurity \citep{csec2017}, and, most recently, the computing competencies of data science \citep{dstf2021}, under the common umbrella of CC2020 \citep{cc2020}. A college that offers degrees across this family is therefore not replicating one program under several names but instantiating several disciplinary specifications that overlap by design in a shared computing core and diverge in their majors. This disciplinary differentiation is the curricular counterpart of the labor market's own segmentation of computing roles, and it is what renders the question of whether a portfolio's programs are differentiated in line with the market both well posed and answerable, since the guidelines establish what each program is meant to emphasize while the market establishes what each corresponding family of roles demands.

A separate body of work, drawn from the measurement tradition in education rather than from computing, bears directly on how the curricular side of such a comparison ought to be read. Research on the \emph{opportunity to learn}, originating in Carroll's model of school learning \citep{carroll1963model} and developed through the large international curriculum studies, distinguishes the \emph{intended} curriculum that appears in official documents from the \emph{implemented} curriculum that is actually taught and the \emph{attained} curriculum that students ultimately acquire \citep{travers1989iea}. The distinction is consequential because the three rarely coincide, and treating the intended curriculum as though it were the attained one inflates any estimate of what graduates know \citep{mcdonnell1995opportunity}; the same concern animates the methodological literature on measuring the content actually delivered rather than its nominal description \citep{porter2002measuring}, in content and in the cognitive depth at which it is pitched, the latter conventionally expressed on the revised Bloom taxonomy and central to the constructive-alignment tradition \citep{anderson2001taxonomy, biggs2011teaching}. A program's published plan of study is an intended curriculum in exactly this sense, and the elective and general-education structures that pad a catalog widen the gap between what is offered and what any one graduate attains. We take this distinction as the conceptual basis for the opportunity--attainment scope model developed below, which reads the curricular supply not as a catalog but on the realized-attainment basis that the opportunity-to-learn tradition prescribes.

A complementary, computational line of work measures that match directly by mining curricular and labor-market text. Early efforts applied keyword and topic-model comparisons to job postings and course catalogs \citep{pirog2024utilising, almgerbi2022systematic, gurcan2019bigdata}; a second wave added supervised classification, named-entity recognition, and taxonomy grounding to extract skills more precisely \citep{karakolis2022bridging, spada2022universities, gnehm2022evaluation, zhang2022skillspan, decorte2023extreme}; and a third has begun to apply large language models to schema-constrained competency extraction \citep{jaiswal2025understanding, zamecnik2024mapping, xu2025course}. A smaller set of studies turns the analysis explicitly toward curriculum design, contrasting the skills taught with the skills sought \citep{foll2018identifying, ahadi2022skills, aljohani2022bridging}. The full trajectory, and the methodological limitations that motivated the reference framework adopted here, are surveyed in \anon{\citet{turaev2026framework}, whose single-program instantiation is reported by \citet{turaev2026pilot}}{prior work, whose single-program instantiation is reported separately (citations withheld for anonymous review)}. What this body of work shares, with few exceptions, is its unit of analysis: it diagnoses one program, or one occupational slice, in isolation.

Common to the more reliable of these efforts is a standardized skills taxonomy that renders the heterogeneous vocabulary of curricula and the market commensurable. The European ESCO classification \citep{esco2021handbook}, the United States O*NET system \citep{onet}, and the Skills Framework for the Information Age \citep{sfia} each supply such a reference vocabulary; we adopt ESCO, for the reasons set out in the framework, as the normalization anchor against which both sides of the comparison are mapped.

The portfolio scale has accordingly remained unexamined. A college of computing offers not a single degree but a family of related programs that share a common core, partition the discipline among their majors, and are intended, collectively, to span the computing labor market; the decisions that govern this arrangement, namely which competencies to guarantee to every graduate, which to localize in a particular program, and how the programs should differ from one another, are made at the level of the portfolio rather than the individual course. A single-program study, however carefully conducted, cannot determine whether a diagnosed gap is idiosyncratic to one program or common to all, whether the shared core carries its weight, or whether the programs are differentiated in a manner that matches the market's own segmentation of computing roles. Answering these questions requires that every program be measured with one and the same instrument, against one and the same labor market, and on a basis that reflects what graduates actually attain rather than what each catalog nominally lists. In the current research, we supply such an analysis, applying a uniform, taxonomy-anchored instrument across an entire undergraduate computing portfolio and reading the curricular supply through the realized-attainment scope model developed in the next sections.

\section{The Alignment Framework}

This study operationalizes the reference framework introduced in \anon{\citet{turaev2026framework} and first instantiated, for a single program, by \citet{turaev2026pilot}}{prior work and first instantiated, for a single program, in its pilot (citations withheld for anonymous review)}; we recapitulate here only what is needed to make the present portfolio-scale analysis self-contained, and we refer the reader to those works for the full formal development. The framework represents every curricular and labor-market requirement as a structured competency record carrying a \emph{domain}, drawn from ten ESCO-aligned computing domains augmented by a transversal-skills category; a \emph{level}, encoding cognitive depth on a five-point ordinal scale derived from the revised Bloom taxonomy \citep{anderson2001taxonomy}; and \emph{evidence}, tying the record to the source passage that produced it, so that every downstream finding remains traceable to the text. Where the pilot populated a seven-slot record by a two-model ensemble, the present work obtains the record by the \emph{grounded} single-model extraction detailed in Section~\ref{sec:method}, which copies each competency verbatim from the source and verifies it against the source, so that the evidence is a grounding guarantee rather than a post-hoc confidence score and the analysis turns on the \emph{domain} and \emph{level} the record carries. Curricular competencies constitute the \emph{supply} inventory and labor-market competencies the \emph{demand} inventory, and the relationship between them is quantified along three dimensions, namely \emph{coverage}, the proportion of demanded competencies for which the curriculum offers a sufficiently similar counterpart; the \emph{cognitive-depth differential}, the signed difference between the level at which a competency is taught and the level at which it is demanded; and \emph{temporal lag}, the delay between a competency's first appearance in the labor market and its subsequent incorporation into the curriculum. The ESCO taxonomy \citep{esco2021handbook} serves throughout as the \emph{normalization anchor} against which both inventories are mapped before they are compared, which renders competencies expressed in different registers and institutional conventions commensurable. Whereas the pilot established the feasibility of this machinery on a single program, the contribution of the present work is to apply it uniformly across an entire college, on a grounded and independently validated extraction, and to add the cognitive and curricular instrumentation that a portfolio-scale, governance-oriented analysis requires.

\section{Study Design and Method}\label{sec:method}

\begin{figure}[t]
  \centering
  \includegraphics[width=\linewidth]{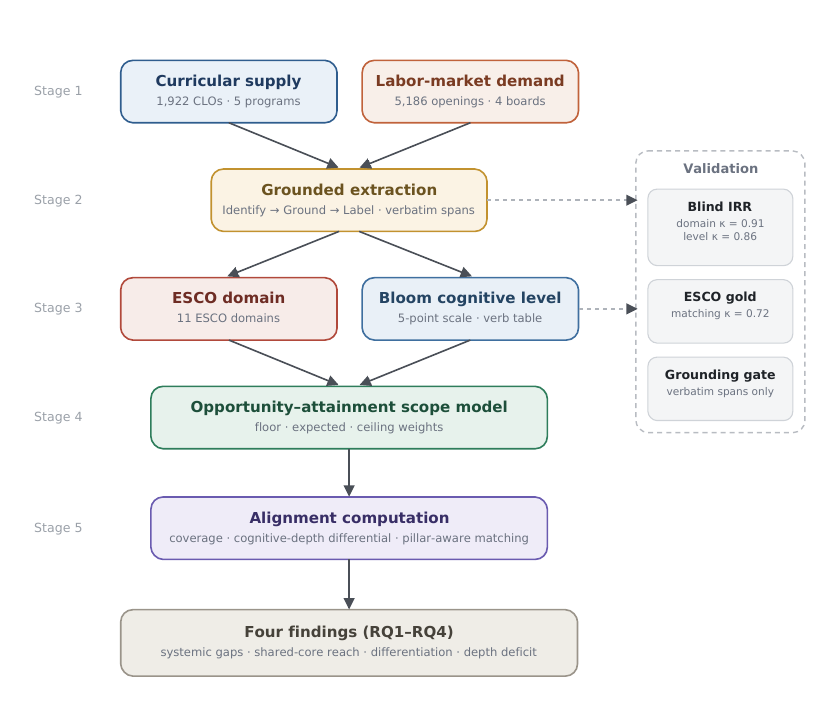}
  \caption{Overview of the portfolio-scale alignment method. The pipeline proceeds in five stages, from the curricular supply and the unified labor-market demand inventories (Stage~1), through the grounded extraction of verbatim competencies (Stage~2) and their anchoring to an ESCO domain and a Bloom cognitive level (Stage~3), to the opportunity--attainment scope model (Stage~4) and the taxonomy-anchored alignment computation (Stage~5), which together yield the four findings. The dashed panel summarizes the independent validation that underwrites the records, namely the blind inter-rater reliability of the extraction, the human-adjudicated ESCO matching gold, and the deterministic grounding gate.}
  \label{fig:method}
\end{figure}

As a case study that applies the portfolio-alignment framework to one institution, the study extends the single-program pilot to the complete undergraduate portfolio of a College of Information Technology at a public research university, comprising five accredited Bachelor of Science programs, namely Computer Science, Computer Engineering, Data Science and Artificial Intelligence, Information Security, and Information Technology. The method proceeds in five stages (Figure~\ref{fig:method}), namely the construction of a curricular supply inventory and a unified labor-market demand inventory; the \emph{grounded} extraction of competencies from each inventory; the anchoring of every competency to an ESCO domain and to a cognitive level on the revised Bloom taxonomy; an opportunity--attainment scope model that reconciles the catalog breadth of each program with the credit-hour and elective constraints under which a graduate realizes it; and the taxonomy-anchored alignment computation that compares the two sides. We describe each stage in turn, and we make the extraction stage, which departs most from the pilot, the subject of the most detailed treatment.

\subsection{The Curricular Supply and the Labor-Market Demand}

The supply inventory is assembled from the complete curricular record of the five programs, rather than from the computing core alone, so that the portfolio is represented as students actually encounter it. Each program organizes its courses into curricular blocks, namely general education, college (computing) requirements, major requirements, major or specialization electives, supporting-requirement groups, and an internship, and the programs share a substantial college-requirement core while differing in their major requirements and electives. Every course contributes its intended learning outcomes, and each \emph{course learning outcome} (CLO) becomes the atomic unit of the supply side, which preserves the granularity at which a curriculum committee actually acts and yields a supply inventory of $1{,}922$ CLOs across the five programs. The decision to treat the CLO as the unit is consequential for what follows: a single outcome corresponds to a single competency, so that the supply inventory neither fragments an outcome into disconnected phrases nor merges several outcomes into one.

The demand inventory is assembled from contemporaneous online job advertisements within the national and regional labor market, filtered by occupation family, namely the group of advertised roles whose discipline corresponds to one of the five programs. Because no single job board offers either complete or unbiased coverage of a regional market, we draw postings from four sources, namely a general aggregator and three regional boards, and we combine them into one corpus. A naive union, however, would overstate demand, since a single opening is frequently cross-posted to several boards and the same role is re-advertised over time, and counting such duplicates would inflate precisely the applied-skill domains that the high-volume boards favor. We therefore \emph{deduplicate} the union to unique openings, collapsing postings that share an employer and a job title (with a description guard for the rare case of an empty employer field) and retaining a single representative per opening together with the list of boards on which it appeared. The union of $6{,}563$ postings reduces in this way to $5{,}186$ unique openings, of which $865$ had been posted more than once, and it is this deduplicated corpus, rather than any single source, that constitutes the demand side of the analysis.

\subsection{Grounded Competency Extraction}

Extracting a structured competency from a course outcome or a job advertisement with a large language model is reliable only to the extent that the model reports what the source says rather than what it expects a plausible source to say. A model asked to summarize the competencies in a passage may paraphrase a requirement into a skill the text does not contain, may supply a governing verb the source omits, or may import competencies that are conventional for the role but absent from the posting, and each such liberty contaminates a downstream coverage or depth measurement in a way that is difficult to detect after the fact. The pilot bounded these risks with a two-model ensemble whose disagreements were flagged for adjudication. To overcome the residual fabrication risk and the cost and complexity that a second extractor entails, the present work adopts a \emph{grounded} extraction procedure built on a single large language model (Claude Opus~4.8) whose every output is verified against the source by deterministic, model-free machinery.

The procedure proceeds in three steps. In the \emph{identify} step, one call to the language model reads the source and returns each competency as a span copied \emph{verbatim} from it, under an instruction to copy character-for-character and to add neither a verb nor an object that the source lacks; in parallel, a deterministic gazetteer tagger marks every occurrence of a curricular-topic or labor-market skill term, supplying a model-independent anchor. In the \emph{ground} step, a grounding gate rejects any returned span that cannot be located in the source, so that a competency that survives is, by construction, present in the text that produced it; across the corpus the gate admits the extractions at full grounding, and no action verb is ever introduced that the source did not contain. In the \emph{label} step, the grammatical \emph{flavor} of each competency, namely whether it states an action the graduate performs or a possession such as knowledge or a credential, and the governing source verb where one exists, are read deterministically from the span. A course learning outcome, being a single self-contained statement, yields exactly one competency, whereas a job advertisement yields several; on the demand side, where postings rarely state a performance verb, the majority of competencies are possessions for which no cognitive level can be honestly assigned, a property we record rather than impute.

\subsection{Taxonomic and Cognitive Anchoring}

Two reference spaces anchor every competency, one for what it concerns and one for the cognitive depth at which it is taught or demanded. For the semantic dimension, each competency is assigned to one of ten ESCO-aligned computing domains augmented by a transversal-skills category. The language model proposes the domain from the competency in context, and a set of deterministic, re-runnable conventions then reconciles the proposal so that boundary cases are treated uniformly across the whole corpus rather than left to the model's case-by-case judgment: foundation mathematics and statistics courses are fixed by course-code rule; outcomes naming \emph{data structures} are placed in Programming and Software Development while those naming an \emph{algorithm} or \emph{recursion} are placed in Algorithms and Computational Theory; cryptographic content remains in Cybersecurity and Ethics even when it names an algorithm; and an outcome concerning the societal or ethical dimension of artificial intelligence is routed to Artificial Intelligence and Data Science when it belongs to an AI course and to Cybersecurity and Ethics otherwise. Separating the extraction of a competency from its typing in this way allows a convention to be revised and the entire inventory re-typed in seconds, without any further call to the model.

For the cognitive dimension, each competency is leveled by its governing verb against a verb-to-level table on the five-point scale of the revised Bloom taxonomy, in which Analyze and Evaluate are collapsed into a single level. The table is a faithful transcription of the computing-specific verb lists published by the ACM Committee for Computing Education in Community Colleges \citep{ccecc2023bloom}, supplemented by the canonical action verbs of \citet{anderson2001taxonomy}, and where the two sources place a verb at different levels we follow the computing-specific assignment, so that, for example, \emph{build} and \emph{produce} are treated as application rather than creation while \emph{design} and \emph{develop} remain creation. A small number of verbs are genuinely polysemous in a way the object resolves: \emph{build} and \emph{produce} are leveled as creation when their object is a whole, novel artifact such as a system, a model, or a program, and as application otherwise. Competencies whose verb falls outside the cognitive taxonomy, principally the affective and dispositional verbs that Bloom's cognitive domain does not cover, and possession competencies that state no verb at all, are recorded as having no assigned level rather than forced onto the scale. The ESCO taxonomy \citep{esco2021handbook}, rebuilt for this study from its official release and partitioned into its knowledge and skill pillars, serves throughout as the normalization anchor against which both inventories are mapped before they are compared.

\subsection{The Opportunity--Attainment Scope Model}\label{sec:scope}

\begin{figure}[t]
  \centering
  \includegraphics[width=\linewidth]{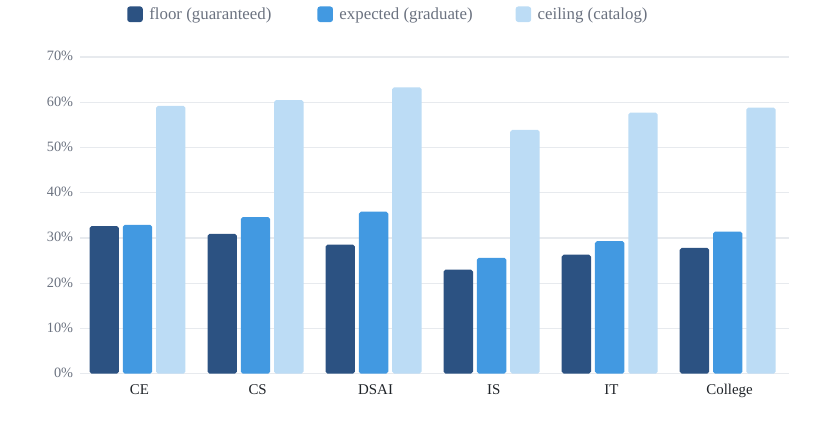}
  \caption{Transversal-skills share of the curricular supply under the floor (guaranteed), expected (graduate), and ceiling (catalog) scopes, by program and pooled. The catalog roughly doubles the transversal share relative to what a graduate is guaranteed (College: $58.8\%$ ceiling against $27.8\%$ floor). Programs: CE (Computer Engineering), CS (Computer Science), DSAI (Data Science and Artificial Intelligence), IS (Information Security), IT (Information Technology).}
  \label{fig:scope}
\end{figure}

A portfolio analysis that counted every catalog course equally would misrepresent what a graduate attains, because the general-education and elective requirements are satisfied by selecting a small number of courses from a much larger menu. We therefore characterize the supply inventory under three scopes: the \emph{floor} counts only the courses every graduate is required to take; the \emph{ceiling} counts the entire catalog; and the \emph{expected} scope weights each elective course by the probability that a graduate completes it under the program's own elective model, namely the number of slots divided by the size of the menu from which they are drawn. The difference between floor and ceiling, which we term the \emph{opportunity--attainment gap}, measures how much of a program's apparent coverage is guaranteed to its graduates rather than left contingent on elective choice. The formal development of the weighting, its normalization, and its verification against each program's prescribed credit-hour load generalize the single-program derivation of the pilot and are recapitulated in Appendix~\ref{app:scope}; Table~\ref{tab:scopemodel} lists the slot counts and menu sizes for the five programs. The effect is substantial: across the portfolio the transversal-skills share of the supply falls from $58.8\%$ at the catalog ceiling to $27.8\%$ at the guaranteed floor, so that nearly half of the apparent general-education breadth is contingent on elective choice rather than assured (Figure~\ref{fig:scope}).

\begin{table}[t]
\caption{Per-program elective model used by the expected scope: the number of mandatory (guaranteed) general-education courses, the general-education elective selection ($s_g$ slots from a pool of $\lvert M_g\rvert$ courses), and the major or specialization elective baskets ($s_g$ of $\lvert M_g\rvert$). Each elective course carries expected weight $\pi = s_g/\lvert M_g\rvert$; mandatory courses carry one.}
\label{tab:scopemodel}
\small
\begin{tabular}{lccl}
\toprule
Program & Mandatory gen-ed & Gen-ed elective $s_g/\lvert M_g\rvert$ & Major/spec.\ elective $s_g/\lvert M_g\rvert$\\
\midrule
Computer Science      & 4 & 3/41 & major 4/10\\
Computer Engineering  & 5 & 2/39 & major 4/12\\
Data Science \& AI    & 4 & 3/41 & specialization 3/5, supporting 3/5\\
Information Security   & 4\textsuperscript{\dag} & 2/40 & major 2/5\\
Information Technology & 4 & 3/40 & major 3/6\\
\bottomrule
\end{tabular}
\\[2pt]
{\footnotesize \textsuperscript{\dag}Plus one mandatory theme satisfied by a choice of one of two equivalent courses, guaranteed at the theme level.}
\end{table}

\subsection{The Alignment Computation}

With both inventories anchored to ESCO, the alignment is computed as a \emph{pillar-aware} match: the object of a competency is matched against ESCO knowledge concepts and the full action clause against ESCO skill and competence concepts, which respects the taxonomy's own division of labor between what is known and what is done. A competency is counted as covering an ESCO concept when their similarity attains a threshold, and per domain we report the supply coverage, the demand coverage, and the demand-minus-supply gap. Because the similarity function and the operating threshold are themselves methodological choices, we do not assume them: the embedding model used for matching, and the distinct model used to refine the domain assignment, were each selected by a task-grounded comparison rather than by reputation, and the coverage threshold is calibrated so that supply coverage attains a reference operating point, with the gap then read at that matched coverage and its sensitivity reported. Matching against the concept label, rather than its longer description, keeps the coverage measure discriminative, since a competency clause embedded against a verbose description saturates the similarity at a low threshold.

\subsection{Reliability and Validation}

The reliability of the analysis is established on the extraction and the matching independently of the demand corpus. The grounded extraction is \emph{faithful} by construction, in that every competency is verified against its source and no governing verb is invented; this is a property of the procedure rather than an estimate. Its \emph{accuracy}, namely whether the assigned domain and level are correct, is assessed by a blind validation in which two independent raters are shown only a competency and its highlighted source and assign the domain and the level themselves, without sighting the model's answer, so that the agreement between raters and the agreement between each rater and the model are both free of the anchoring that a confirmatory ``is this correct'' design would introduce; reliability is reported as Krippendorff's alpha and chance-corrected $\kappa$ coefficients, and the sample deliberately oversamples the hard cases, namely boundary and multi-domain outcomes, multi-verb outcomes, and the verbless possessions, so that accuracy is reported separately for clear and for hard items rather than dominated by trivial ones. The matching is validated by a separate technique-selection gold set of human-adjudicated competency-to-ESCO pairs, against which the candidate matchers are scored, so that the matcher adopted for the analysis is the one that best reproduces human judgment on this task. The agreement, effect-size, multiplicity, and uncertainty measures used throughout are defined in Appendix~\ref{app:stats}.

\subsection{Ethics and Research Integrity}

The blind validation reported in Section~\ref{sec:reliability} involved human raters, who coded competency records under a voluntary and informed protocol, with their responses anonymized before analysis. The labor-market corpus comprises publicly posted job advertisements, which contain no personal data of identifiable individuals, and no posting is reproduced in a form that could identify an employer or applicant. A generative language model was employed as an analytical instrument for the \emph{identify} step of the extraction only, and its every output was verified against the source by the deterministic grounding gate and further subjected to the blind human validation, so that no finding rests on an unverified model assertion, consistent with ACM's policy on the use of generative artificial intelligence in scholarly work.

\section{Reliability of the Extraction}\label{sec:reliability}

The reliability of the structured extraction was established along two complementary axes, namely the agreement between independent human raters on how a competency should be coded and the agreement between the automated extraction and the human consensus, both assessed under a \emph{blind} protocol designed to remove the confirmation bias that a verify-the-system's-answer design would introduce. Two faculty raters independently coded a stratified sample of three hundred competencies, one hundred fifty drawn at random and one hundred fifty drawn deliberately from the hard cases, namely boundary and multi-domain outcomes, outcomes governed by more than one verb, and the verbless possessions that dominate job advertisements, with each rater assigning the ESCO domain and the Bloom level from the competency and its highlighted source alone, without sight of the model's assignment or of the other rater's coding.

Table~\ref{tab:reliability} reports the inter-rater reliability. The raters agree on the eleven-way domain assignment at a Cohen's $\kappa$ of $0.912$ and on the cognitive level at a $\kappa$ of $0.858$, both of which fall in the range conventionally read as almost-perfect agreement, and on the numeric Bloom levels alone the agreement is $93.9\%$ exact with a quadratic-weighted $\kappa$ of $0.877$. The validity judgment returns a high raw agreement of $0.953$ but a modest $\kappa$ of $0.444$, a discrepancy that reflects the well-documented behavior of the $\kappa$ statistic under a skewed marginal rather than unreliable coding: because nearly every extracted unit is in fact a genuine competency, the chance correction that $\kappa$ applies is large, so the prevalence- and bias-adjusted coefficient ($\mathrm{PABAK}=0.907$) and Gwet's $AC_1$ ($0.949$), both robust to such skew, give the faithful picture of near-perfect agreement on validity \citep{feinstein1990paradoxes, gwet2008ac1}.

\begin{table}[t]
\caption{Inter-rater reliability between two independent, blind raters over $300$ competencies. Confidence intervals are $3{,}000$-sample bootstrap. Validity is reported with prevalence-robust coefficients owing to the skewed marginal.}
\label{tab:reliability}
\small
\begin{tabular}{lccccc}
\toprule
Attribute & \% agree & Cohen $\kappa$ & Weighted $\kappa$ & PABAK\,/\,$AC_1$ & $95\%$ CI ($\kappa$)\\
\midrule
Validity (genuine competency) & 0.953 & 0.444 & --- & 0.907\,/\,0.949 & $[0.18, 0.67]$\\
Domain (11 ESCO categories) & 0.930 & 0.912 & --- & 0.924 ($AC_1$) & $[0.88, 0.95]$\\
Bloom level (1--5 + no-level) & 0.887 & 0.858 & --- & 0.865 ($AC_1$) & $[0.81, 0.90]$\\
Bloom level (numeric 1--5) & 0.939 & 0.919 & 0.877 (quad.) & --- & $[0.79, 0.95]$\\
\bottomrule
\end{tabular}
\end{table}

Table~\ref{tab:accuracy} reports the accuracy of the extraction against the human consensus, computed on the cards for which the two raters agree and therefore furnish a defensible ground truth. The extraction is correct on $98\%$ of validity judgments, $99\%$ of domain assignments, and $99\%$ of cognitive levels overall; it is essentially perfect on the random stratum and holds between $96\%$ and $99\%$ on the deliberately hard stratum and on the demand side, where the residual error is concentrated in a small number of boilerplate advertisement phrases that the raters jointly declined to treat as discrete competencies. Because the raters coded blind to the model, this accuracy is free of the anchoring that a confirmatory design would introduce, and the hard-case stratum, on which the accuracy barely declines, furnishes a conservative stress test rather than a best-case estimate.

\begin{table}[t]
\caption{Accuracy of the extraction against the blind human consensus (the cards on which both raters agree), with Wilson $95\%$ intervals reported in the supplementary analysis. Domain and level accuracy are computed on the agreed-domain and agreed-level subsets respectively.}
\label{tab:accuracy}
\small
\begin{tabular}{lcccc}
\toprule
Subset & $n$ & Validity & Domain & Bloom level\\
\midrule
Random (representative) & 150 & 1.00 & 1.00 & 1.00\\
Hard (stress test) & 150 & 0.96 & 0.98 & 0.96\\
Supply (curriculum) & 150 & 1.00 & 0.99 & 1.00\\
Demand (labor market) & 150 & 0.96 & 0.99 & 0.96\\
Overall & 300 & 0.98 & 0.99 & 0.99\\
\bottomrule
\end{tabular}
\end{table}

\begin{figure}[t]
  \centering
  \includegraphics[width=\linewidth]{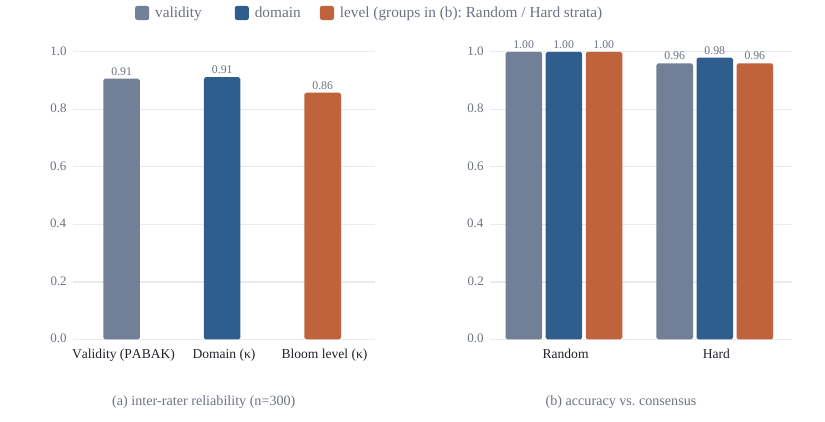}
  \caption{Blind extraction validation. Left: inter-rater reliability on validity (PABAK), domain (Cohen's $\kappa$), and Bloom level (Cohen's $\kappa$); the dashed line marks the almost-perfect threshold. Right: extraction accuracy against the human consensus on the random and hard strata.}
  \label{fig:validation}
\end{figure}

Figure~\ref{fig:validation} summarizes the two together, the blind inter-rater reliability and the model's accuracy across the random and hard strata.

The residual disagreement, modest as it is, was not diffuse but concentrated on two interpretable boundaries. On the domain, the divergences fell almost entirely on the line between the transversal and the technical reading of a competency, and in particular on foundational mathematics courses and on outcomes framed by a communication verb, where one rater coded an outcome such as \emph{communicate the results of an applied linear-algebra computation} by its framing and the other, with the model, by its subject matter. On the cognitive level, the few large gaps arose on multi-verb outcomes, where leveling by the leading verb, the convention the extraction adopts, parts company with leveling by the highest-order verb that the same outcome also contains, as in \emph{define and formulate} or \emph{identify and implement}. In both cases the automated assignment fell within the range of competent human variation rather than outside it, which is the property a reliability analysis seeks to establish; we note the leading-verb convention as a deliberate and reproducible choice rather than a neutral one, and return to its consequences among the limitations.

Together with the procedural faithfulness of the grounded extractor, which admits no span absent from the source and invents no governing verb, these results establish that the competency records on which the portfolio analysis rests are both reliably coded and accurately typed. The validation of the supply-to-demand \emph{matching}, by which competencies are aligned to ESCO concepts, was conducted on the same blind, independent basis. From a matcher-neutral pool of candidate skill-to-concept pairs, the union of the top candidates proposed by every matching technique under consideration, two raters independently judged three hundred pairs as matching or not, agreeing at a Cohen's $\kappa$ of $0.718$, the substantial agreement expected of a semantic-equivalence judgment that is inherently more interpretive than a domain or level assignment; the residual disagreement was itself principled, turning on how much credit a broad, transversal ESCO concept should receive for a specific technical skill. Against the resulting consensus gold of two hundred fifty-eight pairs, the candidate techniques were scored on a held-out split, and a general-purpose sentence embedder (GTE-large) attained the highest $F_1$ of $0.718$, narrowly ahead of a cross-encoder reranker and of the multilingual embedder used by the pilot (BGE-M3); this embedder, which also serves the domain refinement, is therefore adopted for the alignment, and we report the coverage analysis under it with the next-best embedder as a robustness comparison.

\section{Results}

The demand side of the analysis is the unified labor-market corpus, in which the postings collected from a general aggregator and three regional boards are merged and reduced to $5{,}186$ unique openings, from which $103{,}349$ competencies are extracted; the supply side is the curricular census of $1{,}922$ course learning outcomes. We report the four research questions in turn.

\subsection{The content gaps are systemic, not program-specific (RQ1)}

Projecting both inventories onto the eleven ESCO domains and comparing their compositional shares locates where the portfolio's attention diverges from the market's. The two compositions differ substantially (Cram\'er's $V = 0.213$), and the divergence is not uniform but concentrated, as Table~\ref{tab:composition} shows. The market demands competencies in Software Engineering and Project Management at four times the rate at which the curriculum supplies them, in Systems and Infrastructure at three times, and in Web and Mobile Development and in Cybersecurity at roughly two-and-a-half and one-and-a-half times. The curriculum, conversely, over-invests relative to demand in General and Transversal skills and, most strikingly, in Algorithms and Computational Theory, which accounts for $8.7\%$ of the curricular supply against $0.2\%$ of market demand, a representation ratio (a domain's share of demand divided by its share of supply) of $0.02$. The under-supplied domains are precisely the applied and professional ones rather than the foundational ones, and because they are thin in the shared structure that every program inherits, the shortfall is systemic across the portfolio rather than peculiar to any single program, a reading that the program-level analysis of RQ3 makes precise.

\begin{table}[t]
\caption{Domain composition of supply and demand and the demand-to-supply representation ratio (a ratio above one marks under-supply). Cram\'er's $V = 0.213$.}
\label{tab:composition}
\small
\begin{tabular}{lrrrl}
\toprule
Domain & Supply \% & Demand \% & D:S & Status\\
\midrule
Software Engineering \& Project Mgmt & 3.1 & 12.5 & 4.00 & under-supplied\\
Systems \& Infrastructure & 4.9 & 15.3 & 3.11 & under-supplied\\
Web \& Mobile Development & 1.4 & 3.6 & 2.69 & under-supplied\\
Cybersecurity \& Ethics & 6.6 & 9.9 & 1.50 & under-supplied\\
HCI \& Design & 0.9 & 0.9 & 1.01 & aligned\\
Emerging Technologies & 0.8 & 0.8 & 1.01 & aligned\\
Artificial Intelligence \& Data Science & 9.3 & 8.9 & 0.95 & aligned\\
Programming \& Software Development & 3.5 & 3.3 & 0.95 & aligned\\
Computer Architecture \& Hardware & 2.0 & 1.5 & 0.76 & over-supplied\\
General \& Transversal Skills & 58.8 & 43.1 & 0.73 & over-supplied\\
Algorithms \& Computational Theory & 8.7 & 0.2 & 0.02 & over-supplied\\
\bottomrule
\end{tabular}
\end{table}

\begin{figure}[t]
  \centering
  \includegraphics[width=.95\linewidth]{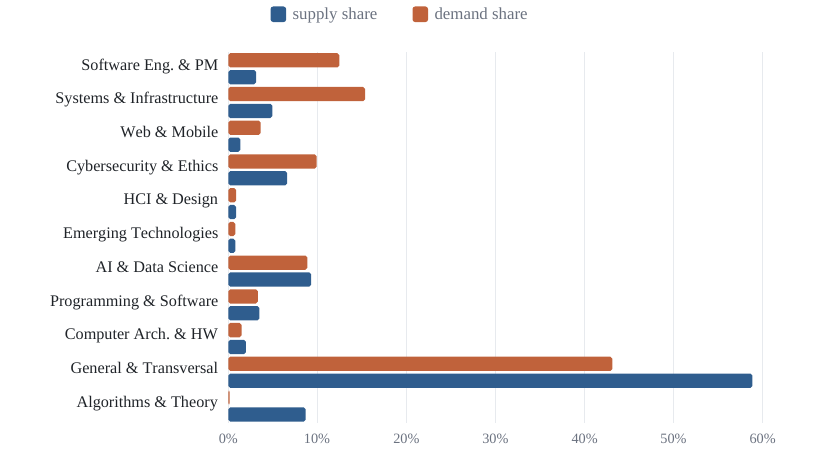}
  \caption{Supply and demand domain composition: each domain's share of the curricular supply against its share of labor-market demand, ordered by the demand-to-supply gap. The applied domains (systems, software engineering, web, security) are demanded far more than supplied; Algorithms and general-education breadth are over-represented.}
  \label{fig:composition}
\end{figure}

\subsection{The shared core is necessary but far from sufficient (RQ2)}

\begin{figure}[t]
  \centering
  \includegraphics[width=\linewidth]{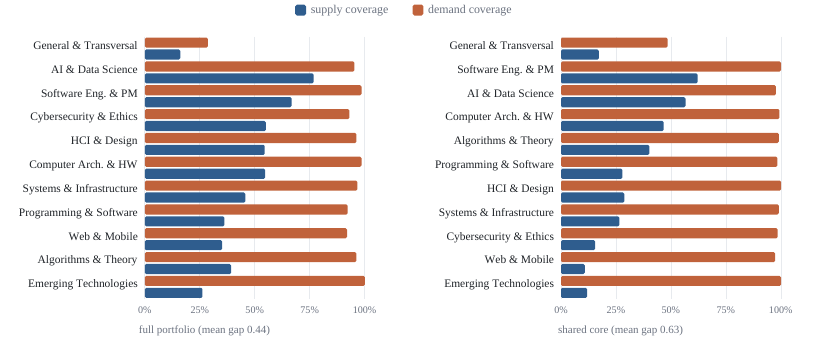}
  \caption{ESCO-concept coverage per domain, supply versus demand, for the full portfolio (left) and the shared college core (right), under the validated matcher. Demand coverage is near-complete; the deficit is the unfilled supply coverage, concentrated in the applied and theory domains and most severe in the shared core.}
  \label{fig:coverage}
\end{figure}

Aligning each side to ESCO and measuring, per domain, the share of demanded concepts the curriculum covers, the full portfolio covers $46\%$ of the demanded competency breadth, leaving a mean content-gap of $0.44$; the deficit is widest in Emerging Technologies ($0.74$), Programming, Algorithms, and Web ($0.57$ each), and Systems ($0.51$), and narrowest in General ($0.18$) and Artificial Intelligence ($0.20$). Restricted to the shared college-requirement core, namely the college-requirement courses common to all five programs that every graduate completes, coverage falls to $0.31$ and the gap widens to $0.63$, with the applied domains all but absent, Emerging Technologies ($0.88$), Web ($0.86$), Cybersecurity ($0.83$), Systems ($0.72$), HCI ($0.71$), and Programming ($0.70$). The shared core thus delivers the foundational and transversal competencies it is designed to guarantee while leaving the applied, market-facing breadth to the program-specific majors. These figures are read at the operating point where supply coverage matches the reference value, and they are stable under the alternative embedder (full-portfolio mean gap $0.48$, shared-core gap $0.64$), so the finding does not depend on the matcher even though the finer per-domain ordering does. Figure~\ref{fig:coverage} shows the per-domain supply and demand coverage for the full portfolio and the shared core.

\subsection{Differentiation in content, homogeneity in shortfall (RQ3)}

\begin{figure}[t]
  \centering
  \includegraphics[width=.62\linewidth]{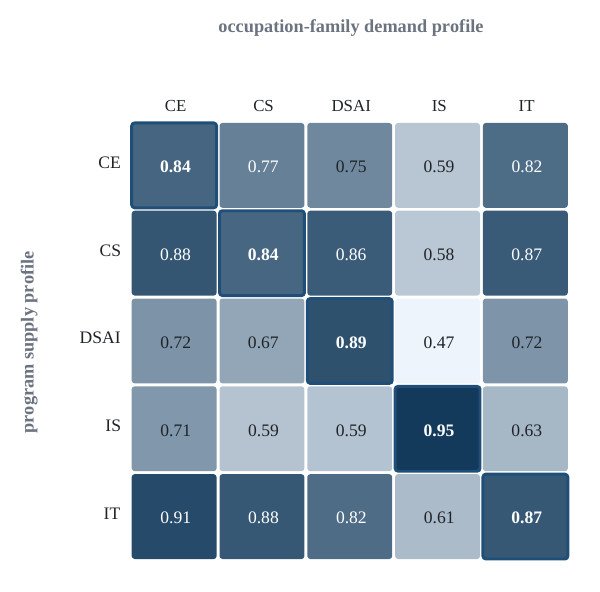}
  \caption{Program supply profiles against occupation-family demand profiles (cosine). Own-family cells (boxed) average $0.88$ against $0.722$ off-diagonal; program-to-program supply similarity falls from $0.835$ on the full curriculum to $0.535$ on the major blocks, locating differentiation in the majors. Codes: CE (Computer Engineering), CS (Computer Science), DSAI (Data Science and Artificial Intelligence), IS (Information Security), IT (Information Technology).}
  \label{fig:progfam}
\end{figure}

The five programs are genuinely differentiated where differentiation is meant to live. On the full realized curriculum their pairwise domain-profile similarity is high ($\overline{\cos} = 0.835$), reflecting the large shared core, but on the major blocks alone it falls to $0.535$, confirming that the programs diverge in their specializations. Each program's supply profile is, moreover, closest to its own occupation family's demand profile more often than not (own-family cosine $0.88$ against $0.722$ off-diagonal), though two of the five, Computer Science and Information Technology, sit nearest the Computer Engineering family, an artifact of the heavily shared computing core. The programs are therefore differentiated in disciplinary content while remaining homogeneous in where they fall short of the market, which is what makes the RQ1 gaps systemic: the under-supplied applied domains are missing from all five programs alike, not localized to one. Figure~\ref{fig:progfam} shows the program-by-family alignment and the drop in program-to-program similarity from the full curriculum to the majors.

\subsection{A portfolio-wide cognitive-depth deficit (RQ4)}

\begin{figure}[t]
  \centering
  \includegraphics[width=.95\linewidth]{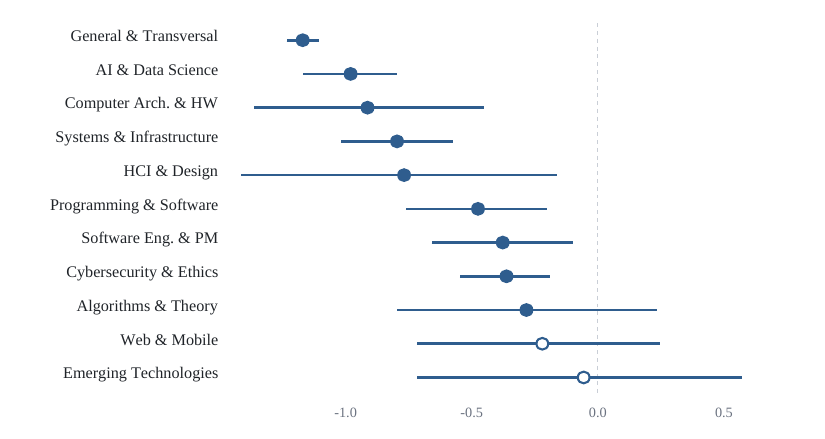}
  \caption{Per-domain Bloom-depth differential (supply minus demand) with $95\%$ confidence intervals; a filled marker denotes significance after false-discovery-rate control. Every domain that admits a comparison is taught below the level demanded, deepest in General, AI, Computer Architecture, and Systems.}
  \label{fig:depthforest}
\end{figure}

\begin{figure}[t]
  \centering
  \includegraphics[width=.85\linewidth]{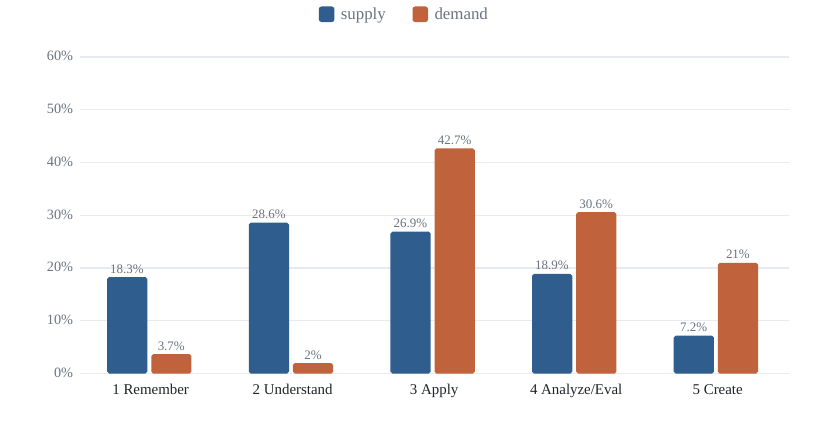}
  \caption{Bloom-level distribution of the curricular supply and the labor-market demand. The curriculum concentrates at the lower levels (Remember through Apply) while the market mass sits at Apply through Create.}
  \label{fig:leveldist}
\end{figure}

Beyond which competencies are taught lies the question of how deeply. Across the portfolio the curriculum is pitched at a mean Bloom level of $2.68$ against the market's $3.63$, a differential of $-0.95$ of a level ($95\%$ CI $[-1.00, -0.89]$; Cliff's $\delta = -0.45$), so the curriculum sits roughly a full cognitive level below market demand. The deficit is significant in nine of the eleven domains and is deepest in General and Transversal skills ($-1.17$), Artificial Intelligence ($-0.98$), Computer Architecture ($-0.91$), and Systems ($-0.80$), and shallowest, to the point of non-significance, in Web and Mobile Development and in Emerging Technologies, where the curriculum already teaches at the applied level the market expects. The depth gap compounds the content gap: the domains the curriculum under-supplies it also, where it does teach them, tends to teach below the level at which the market requires them. Figure~\ref{fig:depthforest} plots the per-domain depth differential with confidence intervals, and Figure~\ref{fig:leveldist} the full Bloom-level distribution on each side.

\section{Discussion and Implications}

Taken together, the four findings describe a curriculum that is broad and shallow where the labor market is applied and deep. The misalignment is at once \emph{structural}, in that the portfolio allocates attention to the eleven domains very differently from the market, leaning toward algorithmic and theoretical foundations and toward general-education breadth while under-serving systems, software engineering, security, and web development, and \emph{cognitive}, in that even where a domain is present it is typically taught roughly a full Bloom level below the depth the market expects. Both forms of misalignment are systemic across the five programs rather than peculiar to any one of them, and they persist on the realized-attainment basis that counts only what a graduate is guaranteed to encounter, which is to say that they are properties of the degrees as they are actually completed rather than artifacts of catalog breadth. We are careful not to over-read the apparent over-investment in foundations, since a labor market advertises for applied competencies far more readily than for the theoretical groundwork it silently assumes, and the extreme case, Algorithms and Computational Theory at a representation ratio of $0.02$, surely reflects in part that postings rarely name the algorithmic competence they presuppose; the robust half of the structural story, namely the divergence between how the curriculum and the market allocate attention across the eleven domains, is the under-service of the applied domains, which postings state explicitly.

For curriculum governance, the systemic character of the gaps carries a direct consequence. Because the five programs fail the market in the same places, the most efficient remedy is not a sequence of program-by-program adjustments but a portfolio-level response, most naturally located in the shared college core that every graduate completes, and the realized-attainment lens sharpens the prescription by distinguishing a competency absent from the curriculum altogether from one present but reachable only through an elective most graduates never take: the former calls for new content, the latter only for promoting an existing elective into the guaranteed core, a far cheaper intervention the catalog view cannot detect. Systems and Infrastructure is the clearest priority, as it is at once the most under-supplied applied domain in guaranteed volume and among the most under-pitched in depth, and because the analysis localizes both which competencies to strengthen and the cognitive level at which they fall short, a committee can act surgically at the level of individual courses rather than through wholesale redesign.

The differentiation result tempers a common intuition. The five programs are genuinely differentiated in disciplinary content, each leaning toward its own occupation family and diverging in the majors where differentiation is meant to live, and this is appropriate and worth preserving; yet differentiation in content does not, on its own, close the market gaps, because those gaps are shared. A college tempted to respond to a perceived misalignment by further specializing its programs would therefore be addressing the wrong margin, since the programs already diverge where the market diverges and converge precisely where they all fall short. The cognitive-depth deficit reinforces the point, for it concerns not which topics appear in a syllabus but the demand at which they are taught, a property that program-level specialization leaves untouched and that can be raised only by changing how competencies are taught and assessed.

Methodologically, the study makes the case that curriculum analytics is most useful to governance when conducted at the scale at which governance decisions are made and on terms that reflect the degree as completed. Three choices proved consequential. Reading the supply on a realized-attainment basis, rather than counting every catalog course equally, corrected the representation ratios in the domains where it mattered most and is a prerequisite for any honest supply-versus-demand comparison. Grounding the extraction so that every competency is verbatim and verified, and subjecting it to a blind, independent human validation rather than treating the language model as an oracle, allowed the reliability of the records to be quantified rather than assumed. And making the threshold-free domain-by-level grid the primary evidence, with the embedding coverage layer reserved for corroboration and itself validated against a human matching gold, kept the headline findings independent of any free parameter. None of these choices is specific to one college: the instrument transfers to any institution that publishes its curricula and any labor market whose advertisements can be collected.

\section{Limitations}

Several limitations bound the study, which is, in form, a case study that applies a general portfolio-alignment framework to the curricula of a single college. It covers one College of Information Technology, so the specific gap profile, though produced by a uniform and reproducible instrument, should be generalized to other institutions only with care; what transfers is the framework rather than the particular findings. The demand corpus is a contemporaneous snapshot of online advertisements from mid-2026 rather than a longitudinal series, so the temporal-responsiveness dimension of the framework cannot be instantiated here, and each competency counts equally, so the demand profile reflects how often a requirement is stated rather than its weighting by seniority or hiring volume. On the method, the cognitive level is read from the governing verb, and on multi-verb outcomes a leading-verb convention is adopted that an equally defensible highest-order-verb reading would level differently; this accounts for the few large rater disagreements on level and is the deliberate, reproducible choice noted in Section~\ref{sec:reliability}. The blind validation used two raters rather than three, and the extraction accuracy is computed against the cases on which they agree, so it is conditioned on a consensus that excludes the genuinely contested items; the residual domain disagreement, in turn, concentrates on the transversal-versus-technical reading of foundational mathematics and communication-framed outcomes. Finally, while the aggregate and shared-core coverage are stable across matching techniques, the finer per-domain coverage ordering is matcher-sensitive, which is why we report the coverage with the human-validated embedder and treat the per-domain ranking as indicative rather than exact, and why the threshold-free representation and depth analyses, which depend on no embedding, carry the principal weight.

\section{Conclusion}

This paper has taken the college, rather than the individual program, as the unit of curriculum--labor market analysis, applying one uniform, taxonomy-anchored instrument across the full undergraduate portfolio of a College of Information Technology and reading the curricular supply on a realized-attainment basis that respects the credit-hour and elective constraints under which degrees are completed. Built on a grounded, blind-validated extraction and a unified labor-market corpus of $5{,}186$ openings, the analysis establishes that the gaps between curriculum and market are systemic rather than program-specific, that the shared college core satisfies only about a third of the demanded competencies, that the programs are well differentiated in disciplinary content yet homogeneous in where they fall short, and that the curriculum is pitched roughly a full Bloom level below the market across the portfolio, most acutely in systems. The misalignment is thus structural and cognitive at once, common to the programs, and concentrated in the applied domains the market names most insistently.

Beyond the particular diagnosis, the contribution is an instrument and a stance: a reproducible, portfolio-scale analysis that measures every program on the same terms, reflects what graduates actually attain, and reports its own reliability, yielding findings a curriculum committee can act on surgically rather than impressionistically. Several extensions would deepen it. A longitudinal demand series paired with dated revisions of the curricular record would make the temporal-responsiveness dimension operational; weighting the demand corpus by seniority, recency, and hiring volume would sharpen the market signal; a finer resolution of the taxonomy below its broad domains would localize gaps within them; and applied across several institutions, the same instrument would turn a single college's audit into comparative evidence about how computing portfolios, as a class, align with the labor markets they serve.

%% ===================== BACK MATTER =====================
\section*{Acknowledgements}
The authors gratefully acknowledge the College of Information Technology and the Research
and Innovation Office at the United Arab Emirates University for their support of the
article processing charge required for open access publication, as well as for the
institutional resources, including computational facilities, library access, and
infrastructural support, that enabled the literature review and the empirical analysis
underlying this work.

\section*{Funding}
No funding was received for conducting this study. The authors did not receive financial
support from any organization for the submitted work.

\section*{Declaration of Competing Interests}
The authors declare that they have no known competing financial interests or personal
relationships that could have influenced the work reported in this paper.

\section*{Author Contributions (CRediT)}
\emph{Sherzod Turaev:} Conceptualization, Methodology, Software, Validation, Formal
Analysis, Investigation, Data Curation, Writing -- Original Draft, Writing -- Review \&
Editing, Visualization, Supervision, Project Administration.
\emph{Saja Aldabet:} Methodology, Investigation, Software, Validation, Data Curation,
Writing -- Review \& Editing.
\emph{Mary John:} Methodology, Investigation, Software, Validation, Data Curation,
Writing -- Review \& Editing.
\emph{Namya Musthafa:} Visualization, Validation, Writing -- Review \& Editing.
\emph{Mamoun Awad:} Conceptualization, Validation, Data Curation, Writing -- Review \&
Editing, Supervision.
\emph{Nazar Zaki:} Conceptualization, Validation, Resources, Writing -- Review \& Editing.
\emph{Khaled Shuaib:} Conceptualization, Validation, Resources, Writing -- Review \&
Editing.

\section*{Data Availability}
The datasets generated and analyzed during the current study, namely the curricular supply
corpus, the deduplicated labor-market demand corpus, and the extracted competency inventories
with their taxonomic and cognitive annotations, are not publicly available at this stage but
are available from the corresponding author on reasonable request.

\bibliographystyle{abbrvnat}
\bibliography{paper2}

\clearpage
\appendix

\section{Derivation and Verification of the Scope Weights}\label{app:scope}

This appendix derives the weights of the opportunity--attainment scope model from program structure and verifies that the expected-scope weights recover each program's prescribed course count and credit-hour load. It generalizes, to the five-program portfolio, the single-program student-path derivation of the pilot.

\subsection{The weighting function}

Let a program comprise requirement groups $G$. A group $g \in G$ offers a menu $M_g$ of distinct courses, of which a graduate completes $s_g$; a required group has $s_g = \lvert M_g\rvert$ and an elective group has $s_g < \lvert M_g\rvert$. Modeling within-group choice as uniform over the menu, the probability that a graduate completes course $c \in M_g$ is $\pi(c) = s_g/\lvert M_g\rvert$, whence the expected-scope weight is $w_{\mathrm{expected}}(c) = \pi(c)$, the floor weight is $w_{\mathrm{floor}}(c) = \mathbf{1}[s_g = \lvert M_g\rvert]$, and the ceiling weight is $w_{\mathrm{ceiling}}(c) = 1$. The uniform choice is the maximum-entropy distribution consistent with the program's structural constraints; where empirical registration data are available, $\pi$ may be replaced by the observed selection frequencies without altering anything that follows. For an ESCO domain $d$ and a scope $\sigma \in \{\mathrm{floor}, \mathrm{expected}, \mathrm{ceiling}\}$, the supply mass and the domain's share are
\begin{equation}
S_\sigma(d) = \sum_{c} w_\sigma(c)\, \mathbf{1}\!\left[\mathrm{dom}(c) = d\right], \qquad \mathrm{share}_\sigma(d) = \frac{S_\sigma(d)}{\sum_{d'} S_\sigma(d')},
\end{equation}
the sum running over all extracted supply competencies, and the \emph{opportunity--attainment band} of a domain, $\mathrm{share}_{\mathrm{ceiling}}(d) - \mathrm{share}_{\mathrm{floor}}(d)$, measures how much of its apparent presence is contingent on elective choice rather than guaranteed.

\subsection{Normalization}

The expected weights are a faithful accounting of what a graduate completes. Summing over the courses of a single group,
\begin{equation}
\sum_{c \in M_g} w_{\mathrm{expected}}(c) = \sum_{c \in M_g} \frac{s_g}{\lvert M_g\rvert} = \lvert M_g\rvert \cdot \frac{s_g}{\lvert M_g\rvert} = s_g,
\end{equation}
so the expected number of courses a graduate completes from group $g$ equals its slot budget, and summing over all groups,
\begin{equation}
\sum_{c} w_{\mathrm{expected}}(c) = \sum_{g \in G} s_g = N_P,
\end{equation}
the total number of courses the program requires. When the courses of a group share a common credit value $\mathrm{CH}_g$, the expected credit-hour load is $\sum_{g \in G} s_g\,\mathrm{CH}_g$, which equals the program's prescribed total. The general-education block makes the identity concrete: every program requires twenty-one credit hours over seven three-credit slots, of which the four mandatory courses (five for Computer Engineering) carry full weight and the elective pool contributes $\sum_{c} \pi(c) = s_g$ further courses, namely three for Computer Science, Data Science and Artificial Intelligence, and Information Technology, and two for Computer Engineering and Information Security. The expected general-education course count is therefore seven, and the expected load twenty-one credit hours, for every program and independently of which electives a graduate selects. The identity depends only on the slot budgets and not on the menu sizes, which is precisely what makes the floor and ceiling assumption-free bounds and the expected scope an unbiased point estimate between them.

\subsection{Mandatory choose-one themes}

One program, Information Security, satisfies a mandatory general-education theme through a choice of one of two equivalent courses. We treat such a theme as a group with $s_g = 1$ and $\lvert M_g\rvert = 2$, so that each alternative carries expected weight $\pi = \tfrac{1}{2}$ and the two together contribute one guaranteed course; the theme is credited to the floor because a graduate certainly completes one of its courses, even though the identity of that course is contingent. Every other group is either fully required ($s_g = \lvert M_g\rvert$) or a genuine elective ($s_g < \lvert M_g\rvert$), and the slot budgets and menu sizes for all five programs are those of Table~\ref{tab:scopemodel}.

\section{Statistical Analysis}\label{app:stats}

Throughout the analysis we report effect sizes as the primary evidence, because the extracted inventories are a near-census of the five curricula and statistical significance is therefore attained trivially at this scale; the significance tests are secondary, and where the labor-market sample is generalized to the market it represents they are read as inferential. This appendix defines the agreement, effect-size, multiplicity, and uncertainty measures used. The omnibus tests we report, namely the Mann--Whitney $U$ test, the Kruskal--Wallis $H$ test, and the $\chi^2$ test of independence, are standard nonparametric procedures and are used in their usual form.

\subsection{Agreement}

Inter-rater agreement is reported as chance-corrected coefficients rather than as raw concordance. For two raters on the nominal domain we use Cohen's unweighted $\kappa$ \citep{cohen1968weighted}, and on the ordinal Bloom scale Cohen's quadratic-weighted $\kappa$,
\begin{equation}
\kappa = 1 - \frac{\sum_{i,j} w_{ij}\, O_{ij}}{\sum_{i,j} w_{ij}\, E_{ij}}, \qquad w_{ij} = \frac{(i-j)^2}{(K-1)^2},
\end{equation}
where $O_{ij}$ and $E_{ij}$ are the observed and chance-expected proportions of rating pairs over the $K = 5$ levels, and the quadratic weights penalize larger level disagreements more heavily than near-misses. Because the marginal distribution of the validity judgment is highly skewed, where nearly every extracted unit is a genuine competency, the Cohen coefficient understates agreement through the well-known prevalence paradox \citep{feinstein1990paradoxes}; we therefore report, for that judgment, the prevalence- and bias-adjusted coefficient $\mathrm{PABAK} = 2 P_o - 1$ and Gwet's $AC_1$ \citep{gwet2008ac1}, both of which are robust to a skewed marginal. The multi-rater nominal agreement underlying the reliability table is summarized by Krippendorff's alpha \citep{krippendorff2004content}, which generalizes the chance-corrected agreement to arbitrary numbers of raters and missing judgments. We read $\kappa$ on the conventional bands: slight up to $0.20$, fair to $0.40$, moderate to $0.60$, substantial to $0.80$, and almost perfect above.

\subsection{Effect sizes}

For a difference in cognitive depth between two groups of competencies we report Cliff's $\delta$ \citep{cliff1993}, a nonparametric dominance measure,
\begin{equation}
\delta = \frac{\#\{(a,b) : a > b\} - \#\{(a,b) : a < b\}}{\lvert A\rvert\,\lvert B\rvert}, \qquad \delta \in [-1, 1],
\end{equation}
where $a$ ranges over group $A$ and $b$ over group $B$; thus $\delta$ is the probability that a competency drawn from $A$ is taught more deeply than one drawn from $B$ minus the reverse, and is robust to the ordinal, non-normal level scale. We interpret $\lvert\delta\rvert$ as negligible below $0.147$, small below $0.33$, medium below $0.474$, and large otherwise. For the association between two categorical variables, such as side and domain, we report Cram\'er's $V$ \citep{cramer1946},
\begin{equation}
V = \sqrt{\frac{\chi^2}{n\,(\min(r, c) - 1)}},
\end{equation}
with $\chi^2$ the test statistic, $n$ the table total, and $r$ and $c$ the numbers of rows and columns, which rescales the $\chi^2$ statistic to $[0, 1]$ (small near $0.1$, medium near $0.3$, large near $0.5$).

\subsection{Multiplicity and uncertainty}

When a test is applied separately across the eleven domains, we control the false-discovery rate by the Benjamini--Hochberg procedure \citep{benjamini1995fdr}: ordering the $p$-values $p_{(1)} \le \cdots \le p_{(m)}$, the adjusted value is
\begin{equation}
q_{(k)} = \min_{\ell \ge k}\, \min\!\left(1,\ \frac{m}{\ell}\, p_{(\ell)}\right),
\end{equation}
and a domain is reported as under-pitched when $q < 0.05$, which bounds the expected proportion of false positives among the flagged domains rather than the familywise error rate. Uncertainty in the portfolio-wide cognitive-depth differential is quantified by a nonparametric bootstrap \citep{efron1993bootstrap}: the matched supply and demand levels are resampled with replacement $B = 3{,}000$ times, the difference of means is recomputed on each resample, and the reported $95\%$ interval is read from the resulting bootstrap distribution.

\subsection{Similarity}

The alignment between a program's supply profile and an occupation family's demand profile (Figure~\ref{fig:progfam}) is the cosine similarity of their eleven-dimensional domain-share vectors $\mathbf{u}$ and $\mathbf{v}$,
\begin{equation}
\cos(\mathbf{u}, \mathbf{v}) = \frac{\mathbf{u} \cdot \mathbf{v}}{\lVert \mathbf{u}\rVert\,\lVert \mathbf{v}\rVert},
\end{equation}
which measures the angle between the two distributions of attention over domains and is invariant to their overall magnitude.

\end{document}